\documentclass[11pt,fleqn,twoside]{article}
\usepackage{amsfonts,amssymb,latexsym}
\makeatletter
\newcommand{\prava}{\footnotesize\it
\begin{flushright}
\begin{minipage}{18cm}
Copyright \copyright 1998 by J. Casahorr\' an 
\end{minipage}
\end{flushright}}

\newcommand{\name}[1]{\begin{flushleft}
                       \LARGE \bf #1
                       \end{flushleft}\vspace{-3mm}}

\newcommand{\Author}[1]{\begin{flushleft}
                       \it #1 \end{flushleft}}

\newcommand{\Adress}[1]{\begin{flushleft}
                       \it #1 \end{flushleft}}

\newcommand{\Date}[1]{\begin{flushleft}
                      \small  \it #1 \end{flushleft}}

\newcommand{\ehkol}{Author \ name}
\newcommand{\ohkol}{Article \ name}
\renewcommand{\@evenhead}{
\hspace*{-3pt}\raisebox{-15pt}[\headheight][0pt]{\vbox{\hbox to \textwidth
{\thepage \hfil \ehkol}\vskip4pt \hrule}}}
\renewcommand{\@oddhead}{
\hspace*{-3pt}\raisebox{-15pt}[\headheight][0pt]{\vbox{\hbox to \textwidth
{\ohkol \hfil \thepage}\vskip4pt\hrule}}}
\renewcommand{\@evenfoot}{}
\renewcommand{\@oddfoot}{}

\newcommand{\be}{\begin{equation}}
\newcommand{\ee}{\end{equation}}
\newcommand{\ba}{\hspace*{-5pt}\begin{array}}
\newcommand{\ea}{\end{array}}

\makeatother

\begin{document}
\setcounter{page}{371}
\thispagestyle{empty}

\renewcommand{\ehkol}{J. Casahorr\'an}
\renewcommand{\ohkol}{Solving Simultaneously Dirac and Ricatti  
Equations}

\begin{flushleft}
\footnotesize \sf
Journal of Nonlinear Mathematical Physics \qquad 1998, V.5, N~4,
\pageref{casahorran-fp}--\pageref{casahorran-lp}.
\hfill {\sc Letter}
\end{flushleft}

\vspace{-5mm}

\renewcommand{\footnoterule}{}
{\renewcommand{\thefootnote}{} \footnote{\prava}}

\name{Solving Simultaneously Dirac and \\
Ricatti  Equations}\label{casahorran-fp}

\Author{J. CASAHORR\'AN }

\Adress{Departamento de F\'{\i}sica, Facultad de Ciencias,
Universidad de Oviedo, \\
Calvo Sotelo s.n., 33007 Oviedo, Spain}

\Date{Received May 15, 1998; Accepted August 19, 1998}

\begin{abstract}
\noindent
We analyse the behaviour of the Dirac equation in $d= 1+1$  with
Lorentz scalar potential. As the system is known to provide a physical
realization of supersymmetric quantum mechanics, we take advantage
of the factorization method in order to enlarge the restricted
class of solvable problems. To be precise, it suf\/f\/ices to integrate a
Ricatti equation
to construct one-parameter families of solvable potentials. To
illustrate the procedure in a simple but relevant context, we resort to a
model which has proved useful in showing the phenomenon of fermion
number fractionalization.
\end{abstract}

\section{Introduction.}

When solving the Dirac equation in $d = 1 + 1$ with  Lorentz scalar
potential, the
underlying supersymmetric structure is crucial. As the system
provides a physical realization of supersymmetric quantum mechanics
(susy qm henceforth), the problem reduces itself to a pair of
Schr\"odinger-like hamiltonians related by means of supersymmetry. In
doing so, both operators share
identical energy spectra unless the Dirac equation allows the existence of a
zero-mode (i.e. a normalizable solution with null energy). Accordingly, we
would like to show how the methods used in susy qm to obtain analytical
solutions of the Schr\"odinger equation can be extended to the Dirac case.
As a relevant example of the aforementioned models one can consider
the problem of the Dirac hamiltonian with a scalar potential
$\phi(x)$ which characterizes a position dependent band gap. In such
a case the Dirac equation serves to describe the behaviour of a
$x$-dependent valence and conduction-band edge of 
semiconductors  in the Brillouin zone~\cite{casahorran_dn}.
On the other hand, the modif\/ied Korteweg-de Vries (mKdV) equation exhibits
a striking relation to the Dirac equation with a time-dependent scalar
potential. If $\phi(x,t)$ represents a solution of the mKdV equation, the
spectrum of the (time-dependent) Dirac operator with the scalar potential
$\phi(x,t)$ is time independent. In other words, the Dirac operators at
dif\/ferent times are unitarily equivalent if the potential itself evolves
according to the mKdV equation~\cite{casahorran_gr}. 

Particular interest acquire however the models in which the Lorentz
potential can be looked upon as
the kink-like static solution derived from a  classical
f\/ield theory. When going to the quantum level, such solutions give rise in
general to new sectors beyond those that are seen in the standard perturbative
regime. These classical solutions
have received  the names of solitary waves, energy lumps, kinks
or solitons. In addition, they appear in any number of
dimensions: the $\phi^4$ kink for $d = 1 + 1$, the Nielsen-Olesen vortex if
$d = 2 + 1$
or the celebrated 't-Hooft-Polyakov monopole in $d = 3 + 1$. (The brave
reader can enjoy the superb description of the subject  contained in the book
of Rajaraman~\cite{casahorran_ra}). Though more massive than the elementary
excitations, the
soliton-like conf\/igurations become stable since an inf\/inite energy barrier
separates them from the ordinary sector. The stability is reinforced by the
existence of a topological conserved charge which does not arise by Noether's
theorem from a well-behaved symmetry of the lagrangian, but takes into
account the large distance behaviour of the f\/ield conf\/iguration
(hence the topological character of the solutions themselves). 

On the other hand,
dramatic ef\/fects appear when a Fermi f\/ield is coupled to the
soliton. Among other things, such models have proved quite useful in
the context of the fermion number fractionalization which has been
observed in certain class of polymers like polyacetylene~\cite{casahorran_ra}.
Describing the 
soliton by a non-dynamical c-number f\/ield, the behaviour of the fermions is
governed in this f\/irst quantized approach by a Dirac operator which
includes the
background provided by the scalar f\/ield. If
one considers interactions enjoying charge conjugation symmetry,  the
states of positive and negative energies match between them.
In addition, the spectrum can exhibit also
self-conjugate zero-energy solutions which as a last resort assume the
responsibility for the fractionalization of the fermion number. These results
are nicely described in topological terms: no matter what the
details of the model, the existence of zero-modes is rigorously predicted
by general arguments concerning dif\/ferential operators.
It may be  worth spelling  that susy qm represents in its own right an
alternative tool
in the analysis of this connection between physics and topology (see
further on).

The organization of the article is as follows. Section 2 takes care of
the Dirac equation in $d = 1 + 1$ with static Lorentz scalar potential.
There we carefully describe the way in which the Ricatti
equation enables us to construct one-parameter families of isospectral Lorentz
scalar potentials for the Dirac case. To illustrate the procedure in a
simple but relevant context, we resort in Section 3 to the problem
consisting in the
analysis of the Dirac equation  on the background provided by the
kink-like conf\/iguration of the $\phi^4$ model. (The f\/irst
appendix includes a capsule introduction to susy qm while the second
one outlines the properties 
of the eigenfunctions associated with the P\"osch-Teller potential).

\section{Supersymmetric structure of the Dirac equation.}

Although there are several applications of susy qm in the context of the Dirac
equation, we discuss only the underlying supersymmetric structure of such
equation in $d = 1 + 1$  with  static Lorentz scalar potential represented
by $f(x)$.
In doing so, we return to the ideas exposed by Cooper {\it et al.}
some years ago \cite{casahorran_co}. The purpose of this section is to show how
we can take advantage of susy qm in order to enlarge in a systematic way the
restricted class of solvable models. To be precise, it is the general
solution of a Ricatti equation that provides us with a one-parameter family of
potentials which serve to extend the f\/irst results derived from $f(x)$.
To begin with, let us give some words about the notation  used in the next
two sections. Space-time coordinates are represented by $x^{\mu}$
$(\mu = 0, 1; x^0 = t, x^1 = x)$. As regards the metric tensor we
have $g_{\mu \nu} = g^{\mu \nu}$,
where $g_{0 0} = - g_{1 1} = 1$ and $g_{\mu \nu} = 0$ otherwise. Moreover
$\partial_{\mu}$ stands for space-time derivatives $\partial / \partial
x^{\mu}$.
When considering a fermion f\/ield~$\psi (x, t)$  moving
in the static Lorentz scalar potential $f(x)$, the behaviour of the system is
governed by the Dirac equation
\begin{equation}
\left[ i \gamma^{\mu} \partial_{\mu} - f(x) \right] \psi(x, t) = 0.
\label{casah_eq:1}
\end{equation}

Letting
\[
\psi (x, t) = \psi (x) \exp  (- i \omega t)
\label{casah_eq:2}
\]
the proper Dirac equation reduces to
\[
\gamma^0 \omega \psi(x) + i \gamma^1 \psi'(x) - f(x) \psi(x) = 0,
\label{casah_eq:3}
\]
where the prime denotes as usual derivative with respect to the
spatial coordinate. Upon choosing now
\[
\gamma^0 = \sigma^1 = \left( \matrix {0 & 1 \cr 1 & 0 \cr} \right),
\qquad \gamma^1 = i \sigma^3 = \left( \matrix {i & 0 \cr 0 & - i \cr}
\right), 
\qquad \psi(x) = \left( {\psi_{-}(x) \atop \psi_{+}(x)} \right),
\]
we get the pair of coupled equations
\[
A \psi_{-}(x) = \omega \psi_{+}(x),
\qquad
A^{\dagger} \psi_{+}(x) = \omega \psi_{-}(x),
\]
where
\[
A = {{d} \over {dx}} + f(x),
\qquad 
A^{\dagger} = -  {{d} \over {dx}} + f(x).
\]

It remains to decouple the equations, i.e.
\[
A^{\dagger} A \psi_{-}(x) = \omega^2 \psi_{-}(x),
\qquad
A A^{\dagger} \psi_{+}(x) = \omega^2 \psi_{+}(x)
\label{casah_eq:12}
\]
which are of the form of susy qm (see appendix A for a brief
introduction to the subject), namely
\begin{equation}
- \psi_{-}{''}(x) + \left[ f^{2}(x) - f'(x) \right] \psi_{-}(x) = \omega^2
\psi_{-}(x),
\label{casah_eq:13}
\end{equation}
\begin{equation}
- \psi_{+}{''}(x) + \left[ f^{2}(x) + f'(x) \right] \psi_{+}(x) = \omega^2
\psi_{+}(x),
\label{casah_eq:14}
\end{equation}
where the Lorentz scalar potential $f(x)$ is just the superpotential
function. To sum up, we substitute the Dirac equation on behalf of the two
Schr\"odinger-like equations of a genuine susy qm model. In such a case,
the components
$\psi_{-}(x)$ and $\psi_{+}(x)$ represent the eigenfunctions of the
hamiltonians
$H_{-} = A^{\dagger} A$ and $H_{+} = A A^{\dagger}$. It is customarily assumed
that whenever the Schr\"odinger equations for potentials $V_{\pm} = f^2(x) \pm
f'(x)$ are solvable, then there always exists a static Lorentz scalar
potential $f(x)$
for which the corresponding  Dirac equation is also exactly solvable.
As expected, the magic of supersymmetry implies that both
hamiltonians $H_{-}$ and $H_{+}$ become solvable simultaneously since the
action of the supercharges  establish the right connection between the
eigenfunctions $\psi_{-}(x)$ and $\psi_{+}(x)$. The  spectrum of energies is
of course degenerate except for the hypothetical zero-mode.  

At this point we can recall the way in which the topological behaviour of the
superpotential $f(x)$ dictates the existence (absence) of zero-modes.
It is customarily assumed
that supersymmetry, a new kind of symmetry which puts together both
bosonic and 
fermionic degrees of freedom, represents  one of the most ambitious
attempts recently made to achieve a trustworthy description of nature.
For better or worse, we
have no experimental evidence for such a novel symmetry being exactly realized
in physics so that one of the most pursued goals in the last two decades has
been to describe a suitable mechanism by means of which supersymmetry shows
up itself spontaneously broken.
As susy qm constitutes an excellent laboratory to analyse the hypothetical
spontaneous breaking phenomenon, Witten  introduced an order parameter
(henceforth the  {\it Witten index} $\Delta$) which in general serves to shed
light on this subject. As a matter of fact the aforementioned
{\it index} $\Delta$ reads~\cite{casahorran_wi}
\[
\Delta (\beta) = \mbox{Tr} \left[ \exp  (- \beta H_{-}) - \exp  (- \beta H_{+})
\right],
\label{casah_eq:15}
\]
where $\beta$ stands for a regularization parameter necessary
when considering susy qm with scattering states.
In doing so, $\Delta (\beta)$ should represent a
measurement of the dif\/ference of $\psi_{-}(x)$ and $\psi_{+}(x)$ modes all
of them with zero-energy since for positive energies duplication occurs
between the two parts of the model. In order not to clutter the
article, we simply point out that
when taking into account the subtleties associated with the counting of
states in the continuous part of the spectrum the {\it Witten index}
$\Delta (\beta)$ satisf\/ies the equation~\cite{casahorran_ac}
\begin{equation}
{{d \Delta (\beta)} \over {d \beta}} = {{1} \over {\sqrt {4 \pi \beta}}}
\left[ f_{+}
\exp \left(- \beta f_{+}^2\right) - f_{-} \exp \left(- \beta
f_{-}^2\right)  \right],
\label{casah_eq:16}
\end{equation}
where $f_{\pm}$ represent the asymptotic values of the
superpotential $f(x)$
as $x \rightarrow \pm \infty$. Although sometimes cumbersome f\/inal
indices are obtained if any of the two limits $f_{-}$ or $f_{+}$
becomes null, it is the case
that whenever we observe a change of sign when going
from $f_{-}$ to $f_{+}$, the integration of (\ref{casah_eq:16}) gives
\[
\Delta (\beta) = \pm \left[ {{1} \over {2}} \Phi \left(f_{+} \sqrt
{\beta}\right) + {{1} \over {2}} \Phi \left(f_{-}\sqrt {\beta}\right) \right]
\label{casah_eq:17}
\]
being $\Phi$ the  Fresnel probability function. Now it
suf\/f\/ices the limit $\beta \rightarrow \infty$ to reveal the
topological roots of the model: the {\it Witten index} amounts to $\pm 1$
so that the asymptotic behaviour of $f(x)$  dictates the existence (absence)
of the non-degenerate zero-mode (see Appendix A).
From a qualitative point of view, this result can also be seen as follows:
a normalizable zero-energy eigenfunction exists for the
superpotentials just exhibiting the right behaviour at inf\/inity.
In summary,
the {\it odd-like} superpotentials  yields a zero-mode
while the {\it even-like} ones are not capable of producing such a
state.

However, the above discussion is only part of the whole story. Once we assume
the existence of a solvable quantum mechanical problem for the pair of
one-dimensional
Schr\"odinger-like equations written in (\ref{casah_eq:13}) and (\ref{casah_eq:14}),
dif\/ferent procedures can be prof\/itably used for
generating one-parameter families of models which share exactly the same
set of eigenvalues. The reasoning becomes much
more accesible when resorting to susy qm in the spirit of the factorization
schemes exposed by Mielnik~\cite{casahorran_mi}.  To
f\/ix the ideas, let us take $f_{+} > 0$ and $f_{-} < 0$ so that a
well-behaved zero-mode exists associated with $H_{-}$. In doing so,
the energy spectrum of $H_{+}$ is positive whereas $H_{-}$ starts
from $E = 0$. Being $\psi_{-0}(x)$ precisely the zero-energy
eigenfunction of $H_{-}$, the 
static Lorentz
scalar potential $f(x)$ appears as the logarithmic derivative of such a
state, i.e.
$f(x) = - \psi{'}_{-0}(x) / \psi_{-0}(x)$. At this stage, we ask ourselves
whether the factorization $H_{+} = A A^{\dagger}$ is unique or can be more
general than usually realized. Writing now a new factorization like
\[
H_{+} = B B^{\dagger}
\label{casah_eq:18}
\]
for
\[
B = {{d} \over {dx}} + F (x), \qquad 
B^{\dagger} = - {{d} \over {dx}} + F (x)
\]
the function $F(x)$ must satisfy the Ricatti-like equation
\begin{equation}
 F{'}(x) +  F^{2} (x)  = V_{+}(x).
\label{casah_eq:21}
\end{equation}

Whenever we know a particular
solution of (\ref{casah_eq:21}) (the potential $f(x)$ itself!), the general
solution can be obtained by putting \cite{casahorran_re}
\[
F (x) = f(x) + \phi (x)
\label{casah_eq:22}
\]
so that
\begin{equation}
\phi{'}(x) + \phi^2(x) + 2 f(x)   \phi(x) = 0.
\label{casah_eq:23}
\end{equation}

Now it proves convenient the introduction of $\rho(x) = 1/ \phi(x)$ to
transform (\ref{casah_eq:23}) into a f\/irst-order dif\/ferential equation like
\[
- \rho{'}(x) + 2 f(x) \rho(x) + 1 = 0
\label{casah_eq:24}
\]
whose general solution corresponds to
\[
\rho(x) = \exp  \left( \int_{-\infty}^{x} 2 f(t) \; dt \right)  \left[
\lambda + \int_{-\infty}^{x} \exp -\left( \int_{-\infty}^{t} 2f(z) \; dz
\right)  dt \right],
\label{casah_eq:25}
\]
where in principle $\lambda$ is a real number. Bearing in mind the
relation between $f(x)$ and the zero-mode $\psi{'}_{-0}(x)$, we get for the
function $F(x)$ the useful expression
\[
F(x) = f(x) + {{\psi_{-0}^{2}(x)} \over {\left( \lambda +
\int_{-\infty}^{x} \psi_{-0}^{2}(t) \; dt \right)}},
\label{casah_eq:26}
\]
whenever the choice for $\lambda$ avoids any singularity along the
real axis. To sum up, we go from the {\it small potential}
$f(x)$ to the {\it large potential} represented by $F(x)$ just by solving
in the standard way a Ricatti dif\/ferential equation. In doing so, we can
take advantage again of the  commutation formula of susy qm to get a family
of  hamiltonians $\tilde{H}$, i.e.
\[
\tilde{H} = B^{\dagger} B
\label{casah_eq:27}
\]
whose close relation with $H_{-}$ can be explained as follows. Except
for the  zero-energy eigenstate, the potential $\tilde{V}(x)$
included in $\tilde{H}$, namely
\[
\tilde{V}(x) = V_{+}(x) - 2   {{d} \over {dx}} \left[  f(x) +
{{\psi_{-0}^{2}(x)} \over {\left(
\lambda +
\int_{-\infty}^{x} \psi_{-0}^{2}(t) \; dt \right)}} \right]
\label{casah_eq:28}
\]
becomes isospectral with respect to the $V_{+}(x)$ itself.
As anticipated in the second section, the zero-modes are found from
f\/irst-order 
dif\/ferential operators so that in our case it suf\/f\/ices to consider
\[
\left( {{d} \over {dx}} + F(x) \right) \tilde{\psi}_0(x) = 0
\label{casah_eq:29}
\]
to obtain
\[
\tilde{\psi}_0(x) = {{\sqrt {\lambda (\lambda + 1)}   \psi_{-0}(x)} \over
{\lambda + \int_{-\infty}^{x} \psi_{-0}^{2}(t) \; dt}}
\label{casah_eq:30}
\]
whenever $\lambda < - 1$ and $\lambda > 0$ in order to guarantee the
right normalization of the state.
In this regards, the  procedure may be understood as a {\it renormalization
of the zero-mode},
leading  to a one-parameter of hamiltonians $\tilde{H}$ isospectral
with the original $H_{-}$. Returning to the Dirac equation itself, the
situation is both
surprising and unexpectedly subtle. First of all, we start from a static
Lorentz scalar potential $f(x)$ which yields solvable
Schr\"odinger-like potentials $V_{\pm}(x) = f^2(x) \pm f'(x)$. Afterwards,
we enlarge in a systematic way the restricted class of such solvable models
just
by going from the {\it small potential} $f(x)$ to the {\it large potential}
$F(x)$ through the
right combination of the  commutation formula of susy qm and the general
theory of Ricatti equations. In addition,
the down components of the spinor remain unchanged whereas the transformation
$\psi_{-}(x) \rightarrow \tilde{\psi}(x)$ must be carried out as far as the
upper components are concerned. In any case, we avoid the dif\/f\/iculties
associated with the handling of the potential $\tilde{V}(x)$ since
the $\tilde{\psi}(x)$ themselves can be obtained
from $\psi_{+}(x)$ by the action of the operator $B^{\dagger}$. 

\section{Illustration: kinks in the presence of Fermi f\/ields.}

To illustrate the above formalism in a simple but physically relevant
scenario, we resort now to the Dirac equation in the background
provided by the kink-like conf\/iguration of the $\phi^4$ model. Such
system has proved quite useful in the context of the so-called
fractionalization of the fermion number which can be observed in
polymers like the polyacetylene. By any standard, the search for
classical solutions represents nowadays one of the most prolif\/ic
areas in quantum f\/ield theory. The interest is mainly motivated by
the belief that such classical conf\/igurations, which of 
course make stationary the action, may shed light on the properties of the
system at issue. In this section we restrict ourselves to non-dissipative
conf\/igurations with f\/inite energy because they constitute the natural
candidates to describe new sectors of the model beyond the perturbative
regime.
On the other hand, dramatic ef\/fects like the fractionalization of the
fermion number appear when a Dirac f\/ield is introduced to the kink itself.
At this stage, it should  be interesting
to review the pioneering  model of Jackiw and Rebbi \cite{casahorran_jr} in the light of
the formalism exposed in the previous section. For reference, we recall that
the  behaviour of the system   is  governed by the  Lagrangian density
\begin{equation}
{\cal L} = {{1} \over {2}} \  \partial_{\mu}\phi\partial^{\mu}\phi -
{{1} \over {2}} \left(\phi^2 - 1\right)^2 +
i\bar{\psi}\gamma^{\mu}\partial_{\mu}\psi -
2\phi\bar{\psi}\psi,
\label{casah_eq:31}
\end{equation}
where as usual $\phi(x)$ and $\psi(x)$ stand for the boson and the
fermion f\/ields  respectively. Notice that for the sake of simplicity the
coupling constants have been absorbed in the f\/ields themselves. Unless
otherwise noted, it suf\/f\/ices here to discuss a f\/irst quantized
approach where the fermion moves in the non-dynamical c-number
f\/ield provided by the static kink $\phi_{k}(x)$. In doing so, the
Dirac equation derived from (\ref{casah_eq:31}) reads 
\[
i\gamma^{\mu}\partial_{\mu}\psi(x,t) - 2\phi_{k}(x)\psi(x,t) = 0
\label{casah_eq:32}
\]
which is of the form written in  (\ref{casah_eq:1}).
In other words, the kink-like prof\/ile represents in this case the static
Lorentz
scalar potential. At this stage, it is customarily assumed the existence of
the well-known kink of the $\phi^4$ model,
namely \cite{casahorran_ra}
$\phi_{k}(x) = \tanh x$
a topological conf\/iguration which smoothly interpolates between
the constant con\-f\/i\-gu\-ra\-tions
$\phi_{-} = - 1$ and $\phi_{+} = 1$.
On comparing now with the general formalism of the
second section, we face a standard Dirac problem with
$f(x) = 2  \tanh x$
whose properties (spectrum and eigenfunctions) are well studied
(see Appendix B).
For reference, we simply point out that in this case we have
\[
V_{-}(x) = 4 - {{6} \over {\cosh^2 x}},
\qquad V_{+}(x) = 4 - {{2} \over {\cosh^2 x}}
\label{casah_eq:36}
\]
while the zero-mode $\psi_{-0}(x)$ reads
\[
\psi_{-0}(x) = {\sqrt {3} \over {2}}   {{1} \over {\cosh^2 x}}.
\label{casah_eq:37}
\]

Now we can write  that
\[
F(x) = 2   \tanh x + {{ 3\cosh^{-4} x} \over { 4\lambda + 3\tanh x -
 \tanh^3 x +2 }}.
\label{casah_eq:38}
\]

As regards the one-parameter family  of potentials $\tilde{V}(x)$ we have
\[
\tilde{V}(x) = 4 - {{6} \over {\cosh^2 x}} - 2   {{d} \over {dx}}
\left[ {{3\cosh^{-4} x} \over {4 \lambda + 3 \tanh x - \tanh^3 x +2 }}
\right]
\label{casah_eq:39}
\]
while the {\it renormalization of the zero-mode} translates into
\[
\tilde{\psi}_{0}(x) = {{2 \sqrt {3} \sqrt {\lambda (\lambda + 1)}\cosh^{-2}
x} \over { 4\lambda + 3\tanh x -
\tanh^3 x +2 }}.
\label{casah_eq:40}
\]

The procedure closes itself  when writing the spinor solutions of
the {\it large potential} $F(x)$   in terms of the
components $\psi_{+}(x)$ and $\tilde{\psi}(x)$ according to the general
method.  

\section{Conclusions.}

In this article we have considered in detail the procedure for obtaining
exact energy
eigenvalues and eigenfunctions for the Dirac equation involving a Lorentz
scalar potential
in $d = 1 + 1$. The generic model is useful in dif\/ferent branches of
physics, ranging from
condensed matter to nonlinear equations. For instance, it serves to study
the fermion
number fractionalization observed in certain class of polymers like
polyacetylene, the
position dependent band gap in semiconductors or the theory of solitons of the
modif\/ied Korteweg-de Vries equation. Taking into account the way in
which the system itself provides us with a realization of
supersymmetric quantum mechanics, we
face ultimately a pair of Schr\"odinger-like hamiltonians. As
expected, the two aforementioned operators share the energy spectra
except for the hypothetical existence of a zero-mode, i.e. a
normalizable solution with $E = 0$. 

In addition, the so-called topological techniques have proved quite useful
in modern physics. When we need to analyze the spectrum of a
well-behaved linear dif\/ferential operator, the
zero-modes are the most crucial ones for applications. To determine the
properties of these modes we can resort to the results contained in
the {\it index theorems}:
no matter what the local details of the model at issue, the existence of
zero-modes depends on the global behaviour of the system (hence the
topological character of the
procedure itself). Taking advantage of the structure associated with susy
qm, it suf\/f\/ices to consider a Ricatti equation in order to
enlarge the restricted class of problems
whose eigenvalues and eigenfunctions can be presented in closed form. To
sum up, whenever we have a pair of solvable Schr\"odinger-like
equations then there also exists a one-parameter
family of Lorentz scalar potentials for which the Dirac equation is exactly
solvable. 

\section*{Appendix A}

In this appendix we present a brief introduction of one-dimensional susy
qm, as tailored
to our needs. In its most simple formulation the model deals with a
time-independent hamiltonian matrix $H_s$ of the form
\[
H_s = \left(\matrix {H_- & 0 \cr 0 & H_+ \cr} \right)
\label{casah_eq:41}
\]
which contains both one-particle hamiltonians $H_-$ and $H_+$. This
operator $H_s$ is part of a graded algebra (superalgebra
in the  jargon of physics) whose closure is achieved by means of the right
mixing of
commutation and anti-commutation relations. At this point, it is
customarily assumed the existence of the so-called superpotential function
$W(x)$ which gives rise to the couple of f\/irst-order dif\/ferential
operators $A$, $A^{\dagger}$ given by
\[
A = {{d} \over {dx}} + W(x),
\qquad A^{\dagger} = - {{d} \over {dx}} + W(x).
\label{casah_eq:43}
\]

Now we can write the supercharges
$Q$, $Q^{\dagger}$, i.e.
\[
Q = \left(\matrix {0 & 0 \cr A & 0 \cr} \right),
\qquad
Q^{\dagger} = \left(\matrix {0 & A^{\dagger} \cr 0 & 0 \cr} \right)
\label{casah_eq:45}
\]
which together with the hamiltonian $H_s$  constitute the
superalgebra of susy qm, namely
\[
H_s = \lbrace{ Q, Q^{\dagger}} \rbrace,
\qquad
\left[H_s, Q \right] = \left[H_s, Q^{\dagger} \right] = 0,
\qquad
\lbrace{ Q, Q} \rbrace = \lbrace{ Q^{\dagger}, Q^{\dagger}} \rbrace = 0.
\label{casah_eq:48}
\]

In more physical terms susy qm can be understood as a matrix hamiltonian
$H_s$ acting on two-component wave-functions $\psi(x)$ like
\[
\psi(x) = \left( {\psi_-(x) \atop \psi_+(x)} \right),
\label{casah_eq:49}
\]
where it should be emphasized that $H_+$ is obtained from $H_-$ just
by reversing the order of the operators $A$ and $A^{\dagger}$. In other words
\[
H_{-} = A^{\dagger} A,
\qquad H_{+} = A A^{\dagger}
\label{casah_eq:51}
\]
 so that the two one-particle hamiltonians $H_{-}$ and $H_{+}$ read
\[
H_{\pm} = - {{d^2} \over {dx^2}} + V_{\pm}(x)
\label{casah_eq:52}
\]
for
\[
V_{\pm}(x) = W^2(x) \pm W'(x),
\label{casah_eq:53}
\]
where the prime denotes derivative with respect to the spatial
coordinate.
It is often the case to refer to the pair of potentials
$V_{-}(x)$ and $V_{+}(x)$ as supersymmetric partners in accordance with the
standard susy qm language. To f\/inish this brief exposition, let us outline the
main properties of our model as far as the standard realization of
susy qm is concerned. First of all, the energy eigenvalues of both $H_{-}$
and $H_{+}$ are positive semi-def\/inite since
the eigenkets $\psi_{-}(x)$ and $\psi_{+}(x)$
verify in fact the relations
\[
\langle \psi_{-} \vert H_{-} \vert \psi_{+} \rangle = {\Vert A \vert \psi_{-}
 \rangle  \Vert}^2,
\qquad
\langle \psi_{+} \vert H_{+} \vert \psi_{-} \rangle = {\Vert A^{\dagger}
\vert \psi_{-}  \rangle  \Vert}^2,
\label{casah_eq:55}
\]
where $\Vert \vert \psi_{\pm} \rangle \Vert$ represents the norm
of the
eigenstate $ \vert \psi_{\pm} \rangle$. As regards the hypothetical zero-energy
solutions $\psi_{-0}(x)$ or $\psi_{+0}(x)$ we have that
\[
A \psi_{-0}(x) = 0,
\qquad 
A^{\dagger} \psi_{+0}(x) = 0.
\label{casah_eq:57}
\]
In doing so, they can be obtained by solving f\/irst-order dif\/ferential
equations so that we write the formal solutions
\begin{equation}
\psi_{\pm 0}(x) \sim \exp \left( \pm \int^x W(y) \; dy \right).
\label{casah_eq:58}
\end{equation}

As any well respected normalizable zero-energy eigenfunction requires boundary
conditions of the type $\psi_{\pm 0} \rightarrow 0$ as $x \rightarrow \pm
\infty$, the meaning of (\ref{casah_eq:58}) is clear: the zero-mode, if
any, must be non-degenerate. The above argument illuminates the framework in
which supersymmetry itself appears either unbroken or spontaneously
broken. To be more specif\/ic, if susy remains exact the energy of
the non-degenerate ground-state is zero while the spontaneous
breaking phenomenon translates into twofold ground-state. On the
other hand, all other energy eigenvalues appear duplicated. When
considering the eigenket $\psi_{-}(x)$ with energy $E_{-}$, i.e.
\[
H_{-} \psi_{-}(x) = E_{-} \psi_{-}(x)
\label{casah_eq:59}
\]
the action of the operator $A$ on the $\psi_{-}(x)$ itself provides
an eigenfunction of the partner hamiltonian $H_{+}$. In other words
\[
H_{+} \left[ A \psi_{-}(x) \right] = A A^{\dagger} A \psi_{-}(x) =
E_{-} \left[ A \psi_{-}(x) \right].
\label{casah_eq:60}
\]

Just by reversing the order of the operators, we start now from the eigenkets
$\psi_{+}(x)$, namely
\[
H_{+} \psi_{+}(x) = E_{+} \psi_{+}(x)
\label{casah_eq:61}
\]
thus obtaining as expected eigenfunctions of $H_{-}$ through the
action of $A^{\dagger}$, i.e.
\[
H_{-} \left[ A^{\dagger} \psi_{+}(x) \right] =  A^{\dagger} A A^{\dagger}
\psi_{+}(x) = E_{+} \left[ A^{\dagger} \psi_{+}(x) \right].
\label{casah_eq:62}
\]

To sum up, except for the hypothetical zero-modes all other energy
eigenvalues appear duplicated when the whole model is studied.
Further on, are
the f\/irst-order dif\/ferential operators $A$, $A^{\dagger}$ that establish
the connection between the two sectors of the system. However, the above
structure exhibits important advantages if we study the problem from
a practical point of view. If we resort for instance to a computational
technique to estimate the energy spectrum, the original hamiltonian
can be substituted on behalf of the isospectral one to improve the
properties of convergence. The same argument can be applied of course
when considering variational methods.  

\section*{Appendix B}

For reference, we outline in this appendix the mean features concerning the
P\"osch-Teller potentials which arise in the analysis of the Dirac equation
on non-trivial backgrounds like the one associated with the kink of the
$\phi^4$ model. The problem itself can
be elegantly solved in a Lie-algebra framework according to the technique
described by Frank and Wolf~\cite{casahorran_fw}, whose notation will be used in the
following.
To start from scratch, let us consider the P\"osch-Teller hamiltonian
\begin{equation}
H_{PT} = - {{1} \over {2}} {{d^2} \over {dx^2}} + {{c} \over {\cosh^2 x}} +
{{s}\over {\sinh^2 x}},
\label{casah_eq:63}
\end{equation}
where we choose a system of units in which $\hbar =  m = 1$. Now
it proves convenient the introduction of the parameters $k_{i}$ $(i = 1, 2)$
given by
\[
c = -  {{1} \over {2}} \left[ (2 k_{1} - 1 )^2 - {{1} \over {4}} \right],
\qquad s =   {{1} \over {2}} \left[ (2 k_{2} - 1 )^2 - {{1} \over {4}} \right]
\label{casah_eq:65}
\]
which in turn enable us to write the energies $E_{k}$ of the bound
states associated with the P\"osch-Teller potential of (\ref{casah_eq:63}), namely
\begin{equation}
E_{k} = - {{ (2 k - 1)^2} \over {2}},
\label{casah_eq:66}
\end{equation}
where $k = k_{\min}, k_{\min - 1},\ldots > 1 / 2$ with $k_{\min} =
k_{1} - k_{2}$.
Letting $k$ to be of the form
\[
k = {{1 + i \kappa} \over {2}}   \kappa \ge 0
\label{casah_eq:67}
\]
the expression of $E_{k}$ in (\ref{casah_eq:66}) also takes care of the
continuous
part of the spectrum. It remains to identify the behaviour of the
wave-functions themselves which, as expected, depends on the values of
the singularity  parameter $s$ in (\ref{casah_eq:63}). At f\/irst glance the
situation is as follows: 

i) For the bound states we have
\[
\Psi_{k}^{(k_{1}, k_{2})} \sim (\cosh x)^{- 2 k_{1} + 3 / 2} (\sinh x)^{
2 k_{2} - 1 / 2}
  F \left[ \alpha, \beta, \gamma; z \right],
\label{casah_eq:68}
\]
where $ F \left[ \alpha, \beta, \gamma; z \right]$ stand for the
$_{2}F_{1}$ Gauss hypergeometric function with the iden\-ti\-f\/i\-ca\-tions
\[
\alpha = - k_{1} + k_{2} + k,
\qquad \beta = - k_{1} + k_{2} - k + 1,
\qquad \gamma = 2 k_{2},
\qquad z = - \sinh^2 x.
\label{casah_eq:72}
\]

ii) When considering the scattering states we f\/ind
\[
\Psi_{k}^{(k_{1}, k_{2})} \sim (\cosh x)^{ 2 k_{1} - 1 / 2}  (\sinh x)^{ 2
k_{2} - 1 / 2} F \left[ \alpha, \beta, \gamma; z \right]
\label{casah_eq:73}
\]
now with
\[
\alpha =  k_{1} + k_{2} - k,
\qquad \beta =  k_{1} + k_{2} + k - 1,
\qquad \gamma = 2 k_{2},
\qquad z = - \sinh^2 x.
\label{casah_eq:77}
\]

With regards to singular potentials like the ones written in
(\ref{casah_eq:63}), the careful analysis of the problem can be carried out
in accordance with the subtle mathematical tools exposed by Reed and
Simon \cite{casahorran_rs}. As far as our model 
concerns, it
might seem plausible to infer the existence of three well dif\/ferent regions
depending on the values of the  parameter $s$, namely: 

\begin{enumerate}

\item[i)] If $s < - 1 / 8$, the model itself becomes physically
unrealistic since the spectrum is not bounded from below. 

\item[ii)] Whenever $- 1 / 8 < s < 3 / 8$, the singularity is not
strong enough to make the wave-functions vanish at the origin.
However, it proves necessary to def\/ine {\it one-parameter family of
self-adjoint extensions} so that a bounded from below spectrum
appears once a proper domain is specif\/ied. In more physical terms,
the wave-functions pass across the singularity point and the model
extends itself along the real axis. 

\item[iii)] Finally, if $s > 3 / 8$ the singularity acts as an
impenetrable barrier thus dividing the space into two independent
regions, i.e. $x > 0$ and $x < 0$, forcing the wave-functions to
vanish at the origin. 
\end{enumerate}

The above results shed light on the family of susy qm models governed by
superpotentials like
\begin{equation}
W(x) = \ell   \tanh x, \qquad   \ell = 1, 2,\ldots
\label{casah_eq:78}
\end{equation}
which give rise to
\[
H_{-} = - {{d^2} \over {dx^2}} + \ell^2 - {{\ell ( \ell + 1)} \over
{\cosh^2 x}},
\qquad
H_{+} = - {{d^2} \over {dx^2}} + \ell^2 - {{\ell ( \ell - 1)} \over
{\cosh^2 x}}.
\label{casah_eq:80}
\]

In accordance with the previous results we have for $H_{-}$
\[
k_{1} = {{\ell} \over {2}} + {{3} \over {4}},
\qquad k_{1} = - {{\ell} \over {2}} + {{1} \over {4}}
\label{casah_eq:82}
\]
together with
\[
k_{2} = {{3} \over {4}},
\qquad k_{2} = {{1} \over {4}}
\label{casah_eq:84}
\]
so that the discrete energy spectrum reads
\[
E_{-j} = \ell^2 - (\ell - j )^2, \qquad j = 0, 1,\ldots, \ell - 1
\label{casah_eq:85}
\]
while the scattering states start at $E = \ell^2$. With regards to
$H_{+}$ we f\/ind
\[
k_{1} = {{\ell} \over {2}} + {{1} \over {4}},
\qquad k_{1} = - {{\ell} \over {2}} + {{3} \over {4}}
\label{casah_eq:87}
\]
and again
\[
k_{2} = {{3} \over {4}},
\qquad 
k_{2} = {{1} \over {4}}
\label{casah_eq:89}
\]
with energies
\[
E_{+j} = \ell^2 - (\ell - j )^2, \qquad j = 1,\ldots, \ell - 1.
\label{casah_eq:90}
\]

The dif\/ference between both discrete spectra
reduces itself to the zero-mode of $H_{-}$ which adopts the form
\[
\psi_{-0} (x) \sim P_{\ell}^{\ell} (\tanh x),
\label{casah_eq:91}
\]
where the $P_{\ell}^{\ell} (\tanh x)$ represent  associated
Legendre polynomials. Bearing in mind that
\[
P_{\ell}^{\ell} (\tanh x) \sim \cosh^{- \ell} x
\label{casah_eq:92}
\]
the logarithmic derivative of the zero-mode itself provides us with
the superpotential functions written in (\ref{casah_eq:78}), i.e.
\[
W(x) = - {{P{'}_{\ell}^{\ell} (\tanh x)} \over {P_{\ell}^{\ell} (\tanh x)}}.
\label{casah_eq:93}
\]

In this case, being $s = 0$, we get rid of the dif\/f\/iculties
associated with the singular behaviour of the potential. Among other
things, the system extends along the real line and the expected
pattern of even and odd wave-functions
(notice that we are dealing with even potential) appears in the end.

\label{casahorran-lp}

\end{document}